# Rainfall-induced Mass Movement as Self-organization Process

Zhengjing Ma[1,2], Gang Mei[1,2*], Nengxiong Xu[1,2], Yongshuang Zhang[1,2], Jianbing Peng[1,2,3]

[1]School of Engineering and Technology, China University of Geosciences (Beijing)
[2]Institute of Geosafety, China University of Geosciences (Beijing)
[3]School of Geological Engineering and Geomatics, Chang'an University
Email: gang.mei@cugb.edu.cn

# Abstract

Self-organizing processes shape Earth's surface, creating complex patterns from simple rules in most landforms. Rainfall-induced mass movements dramatically reshape landscapes through rapid sediment transfer, but whether they self-organize remains unknown. Here we decode their organizational principles by treating spatial changes in scar geometries as fingerprints of the movement process. In 65,936 scars worldwide, we discovered three geometric signals from width, sinuosity and curvature converge on shared patterns and identify a slow-to-fast hierarchy characteristic of self-organizing landforms: long-range correlations show width retaining spatial memory while curvature decorrelates quickly; power spectra quantify a '4-3-2' hierarchy (width-sinuosity-curvature) in scaling exponents; and information flow confirms a top-down organization (width→sinuosity→curvature). Although entropy increases toward finer scales, phase-space reconstructions settle on low-dimensional attractors, revealing hidden order. Together, the evidence shows that width establishes flow corridors through slow dynamics, sinuosity mediates momentum and gravity by intermediate adjustments, and curvature responds rapidly to the terrain. We also developed a model based on simple terrain–inertia trade-offs, demonstrating how mass movements maintain large-scale coherence while flexibly navigating obstacles, potentially extending run-out distances. This organizing rule offers a fundamental mechanism for predicting the destructive reach of mass movements, which are intensifying in our warming, wetter world.

# 1 Introduction

Self-organization processes are ubiquitous in landscape evolution, from evenly spaced valleys to ripples of wind-blown sand with uniform spacing, and from river meandering to tidal-dominated delta distributaries[1,2,3,4,5,6,7]. These processes, encoded in morphological and geometric signatures across Earth's surface, show scale-invariant properties, power-law distributions, and fractal characteristics that transcend specific environmental conditions[8,9,10,11,12]. Laboratory and numerical studies explain these order through nested feedbacks: rapid, small-scale processes interact, larger structures emerge, and, once in place, those structures steer the smaller ones[13,14,15,16]. These internal dynamics, where feedbacks between topography, erosion, and sediment transport operate independently of external perturbations, are increasingly recognized as fundamental organizing principles across Earth's erosional landscapes[17,18]. Systems organized through this feedback typically operate near a critical point where sufficient order maintains form. At the same time, flexibility permits continual adjustments that optimize energy dissipation across the landscape, such as chute cutoffs in meandering rivers and rill formation on hillslopes[19,20]. The persistence of these organizational principles extends even across geological timescales, where landscapes maintain states of dynamic disequilibrium through ongoing geometric reorganization, such as river basin adjustments that can prevent equilibrium for hundreds of millions of years[21].

However, what about one of Earth's most rapid reshaping events? Self-organization has rarely been noticed in rainfall-induced mass movement. These brief and devastating natural hazards are becoming more frequent with intensifying precipitation[22]. A single storm can transfer the bulk of a mountain's sediment budget in minutes, leaving only aftermath scars to interpret, making any hidden organization hard to detect. Do such fleeting processes have organizational rules similar to longer-lived landforms? Evidence suggests they might. Mass movement scars have shown some universal patterns. For example, area-to-volume relationships follow power laws, and shape characteristics remain remarkably uniform across different rock types and slope[23,24]. Furthermore, topological features from scars encode regularities that machine-learning models can use to classify failure modes; yet, such signatures remain abstract and lack a mechanistic explanation[25]. Indeed, granular physics studies have observed that dense-to-dilute transitions during runout produce local chaotic divergence, a signature of self-organized criticality; however, field evidence remains limited[26]. We therefore ask: do rainfall-triggered mass movements contain the same organizational principles as other landforms? If so, how do these principles emerge, and do they leave their signatures in the landscape?

Here we decode the organization rule of mass movement through three geometric signatures: width captures lateral spreading and flow confinement, sinuosity quantifies meandering paths that dissipate energy, and curvature records instantaneous adjustments to terrain obstacles. By treating their spatial changes as dynamic fingerprints, we analyze 65,936 rainfall-induced scars from 26 global inventories and discover that mass movements consistently organize into a limited set of geometric archetypes, revealing universal patterns that persist across diverse settings. We suggest that these patterns appear to arise from a slow-to-fast hierarchy characteristic of self-organizing systems, based on quantitative analysis of spatial signals: long range correlation analysis shows width maintaining long-range persistence while curvature rapidly decorrelates; spectral analysis reveals a remarkable '4-3-2' scaling hierarchy (width-sinuosity-curvature); and information flow confirms asymmetric top-down control (width→sinuosity→curvature). Entropy metrics reveal an increase in disorder at finer scales, while phase-space analysis demonstrates that these dynamics remain bounded on low-dimensional attractors. **Our findings suggest that mass movements self-organize through the similar rules as rivers and dunes**: slow-evolving width defines flow corridors that constrain intermediate sinuosity adjustments and rapid curvature responses. This hierarchical organization enables continuous self-adjustment, where a large-scale order maintains a coherent flow while local grain-scale instabilities provide flexibility to navigate terrain irregularities, thereby maximizing runout distance and optimizing energy dissipation across the landscape. To validate this, we have also developed a conceptual terrain–inertia trade-off model that reproduces the observed flow paths, demonstrating the emergence of mass-movement dynamics from simple rules.

# 2 Results

## 2.1 Geometric fingerprints reveal hidden universal patterns

We first investigated how mass movement dynamics are encoded in scar geometric signatures and the extent to which these movements exhibit ordered spatial organization. From 65,936 scars across 26 event inventories (Fig.1a), we extracted along-track spatial series (termed profiles or signals) for width ($W$, lateral extent), sinuosity ($S$, path deviation), and curvature ($C$, directional change) (Fig.1b and Extended Data Fig.1). From these profiles, we derived geometric attributes quantifying trends, variability, symmetry, and oscillatory behavior (Fig.1c, Supplementary Figs.10-12).

We found remarkably consistent distributions in kernel density estimates across all inventories, pointing to underlying geometric rules (Fig.1d). Width profiles are typically symmetric and uniform, with centroids clustering near the midpoint (0.50 ± 0.025, mean ± s.d.), and high bilateral consistency (symmetry: 0.8 ~ 1.0), suggesting that significant width adjustments concentrate within the main transport section. Sinuosity profiles show no overall directional trends (slope −0.0000 ± 0.0032) but display regular oscillations every 4 ~ 8% of path length, indicating systematic corrections balance directional deviations. Curvature profiles show the most significant local variability, yet their local maxima and minima counts converge within a narrow range (9.0 ± 2.0). Turn-asymmetry distributions consistently form a distinctive triple-peak pattern (representing left-biased, symmetric, and right-biased turn sequences), implying that mass movements operate with a conserved 'budget' of directional change (Supplementary Figs.13-15).

To quantify this observed universality, we applied manifold learning (Uniform Manifold Approximation and Projection, UMAP) to the derived attributes for each geometric signature, capturing distinct archetypes (Fig.1e, Extended Data Fig.2, Supplementary Figs.16-18): Width archetypes include (C0) constricted/tapering profiles, (C1) highly symmetric and uniform profiles, and (C2) profiles with pronounced oscillations. Sinuosity archetypes range from (C0) low and stable to (C1) trend-dominated, asymmetric, and (C2) highly oscillatory patterns. Curvature archetypes comprise (C0) frequent, sharp turns, (C1) trend-dominated and asymmetric turn sequences, and (C2) more symmetric profiles with fewer, broader turns. The consistent emergence of low-dimensional manifold structures of scars across numerous inventories (mean silhouette scores: Width = 0.598, Sinuosity = 0.421, Curvature = 0.453, with Coefficients of Variation (CV) $\ll$ 0.112; mean centroid similarity: Width = 0.955, Sinuosity = 0.993, Curvature = 0.899, with CV $\ll$ 0.267), indicates that mass movements converge toward a limited set of planform characteristics, resulting in a shared structural organization.

We next implemented a Voronoi-based analysis of mass movement scar 'skeletons' as complementary validation, focusing on geometric configurations including longest sliding paths, branches, junctions, and sharp turns. We observed universal geometric patterns (Extended Data Fig.3). (1) Longest runout path lengths consistently exhibited heavy-tailed distributions with power-law tails (mean fitted $a \approx 3.3$). (2) Branch points, normalized by path length, have an upper bound at 0.4, suggesting limits on branching complexity. (3) Sharp turns predominantly occurred with deviation angles between 140° and 160° (mean 144.6° $\pm$ 0.07°), and showed close spacing (mean normalized spacing 0.2632 $\pm$ 0.1667), consistent with the oscillatory behavior observed in planform signals, where inflection point counts systematically increased from width (3.27 $\pm$ 2.60) to sinuosity (7.34 $\pm$ 1.96) to curvature (14 $\pm$ 6). (4) We found a positive correlation between sharp turn count and longest path length (log-log slope $\approx$ 0.48), indicating that longer scars accumulate more turns (Supplementary Fig.19). For high-angle turns ($\gg$ 155.35°), width variability was greater within frequent turn zones (CV = 1.183) than for isolated high-angle turns (CV = 0.682), suggesting that successive directional changes trigger more pronounced local width fluctuations. We conclude that mass movements are shaped by localized adjustments, enabling frequent terrain adaptations while preserving global path efficiency and coherence.

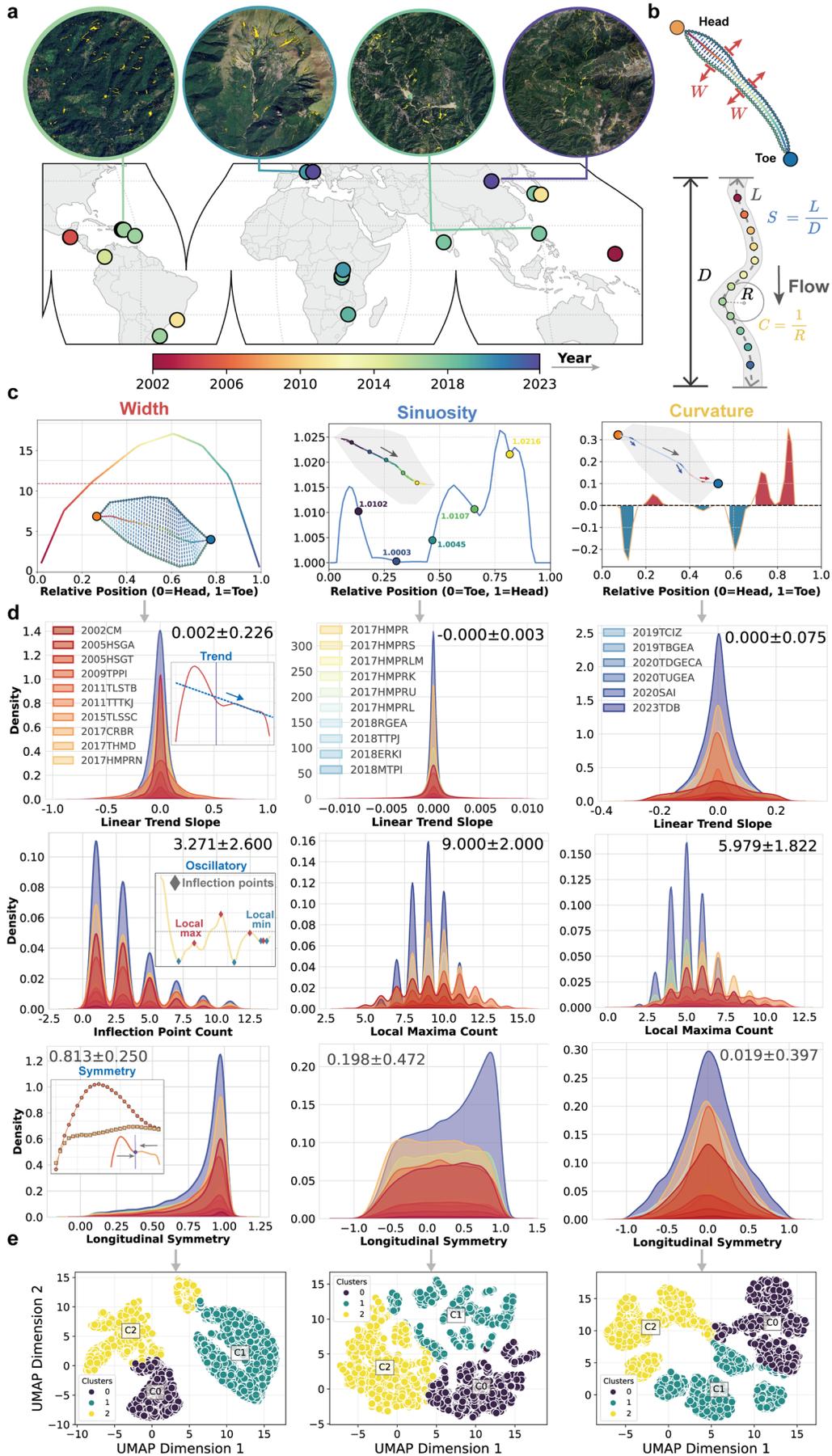
5

**Fig.1: Universal geometric patterns in rainfall-induced mass movement scars. a,** Global map of 26 analyzed mass movement inventories (2002 ~ 2023, color-coded by year) with example satellite imagery of scar networks. **b,** Schematic defining key geometric signatures: width ($W$), sinuosity ($S$), and curvature ($C$) along a mass movement scar. **c,** Representative along-track profiles (normalized position: 0 = head, 1 = toe) for width, sinuosity, and curvature. **d,** Stacked kernel density estimates (KDEs) for key geometric attributes (linear trend slope; inflection / local maxima count; longitudinal symmetry) derived from 65,936 mass movement scars. Consistent distributions across diverse inventories (see legends) highlight universal statistical characteristics. Mean $\pm$ s.d. values for each attribute are indicated above the respective plots. **e,** Uniform Manifold Approximation and Projection (UMAP) of feature sets derived from width (left), sinuosity (center), and curvature (right) reveals three distinct archetypal clusters (C0, C1, C2).

## 2.2 Apparent randomness bounded by underlying constraints

The universal geometric patterns raise a question: what underlying dynamical processes generate such consistent yet locally variable scar geometries? This leads us to explore deterministic chaos, where simple rules produce seemingly unpredictable behavior while maintaining statistical regularity[27, 28].

We began by embedding spatial series of width, sinuosity, and curvature into phase space using delay-coordinate reconstructions, and observed distinct low-dimensional attractors: width profiles traced broad loops, sinuosity profiles formed intermediate coils, and curvature profiles collapsed into dense spirals (Fig.2a,b, Supplementary Fig.20). To quantify these shapes we computed the correlation dimension $D$, obtained from the scaling of the correlation sum $C(r) \propto r^D$. Across embedding dimensions $d = 2$ ~ 15, the estimated $D$ remained well below $d$, confirming low-dimensional deterministic dynamics that has also been observed in meandering rivers (Fig.2c,d, Supplementary Figs.21, 22). Width shows the greatest geometric freedom ($D$ from 1.05 to 0.41 as $d$ increased), sinuosity occupies an intermediate range (0.82 → 0.27), and curvature is tightly constrained (0.85 → 0.09). We interpret this ordering as evidence that width is influenced by external boundary conditions, allowing many geometric configurations, whereas curvature records internally governed, rule-driven adjustments.

We further examined the dynamics underlying the three geometric signatures by measuring Lyapunov exponents. A consistent ordering of exponents ($\lambda_1 > \lambda_2 > \lambda_3 > \lambda_4$) emerged (Supplementary Figs.23, 24). $\lambda_1$ was consistently positive (width: 0.267 $\pm$ 0.385, sinuosity: 0.321 $\pm$ 0.722, curvature: 0.601 $\pm$ 0.957), demonstrating sensitive dependence on initial conditions, enabling transient disturbances to be amplified into disproportionately large differences. $\lambda_4$ was persistently negative (width: -0.411 $\pm$ 0.379, sinuosity: -0.334 $\pm$ 0.382, curvature: -0.528 $\pm$ 0.531), signifying strong contraction in phase space and indicating fundamental constraints governing mass movement, such as gravity-driven downhill motion. The combination of a positive $\lambda_1$ and a negative sum of all exponents ($\sum \lambda_i < 0$) suggests that the mass movement dynamics are both chaotic and dissipative. Principal Component Analysis (PCA) for the Lyapunov exponents confirmed this interpretation, with system dynamics captured by a two-dimensional manifold (67% of total variance). The primary component (PC1) combines divergence ($\lambda_1$) with contraction ($\lambda_4$), while the secondary component (PC2) reflects intermediate dissipative rates ($\lambda_2$, $\lambda_3$), supports this chaotic and dissipative nature (Supplementary Fig.25).

To determine whether chaotic behavior represents a fundamental property of mass movements, we tested whether these dynamical properties systematically change with scar size (small, medium, and large, classified according to power-law size distribution) (Fig.2e, Supplementary Fig.8). We found that the Lyapunov exponent ordering remained consistent across all size classes. $\lambda_1$ showed minimal variation: curvature (0.107, 0.122, 0.100 for small, medium, large scars respectively), sinuosity (0.107, 0.123, 0.099), and width (0.062, 0.073, 0.060). $\lambda_3$ and $\lambda_4$ converged on consistent negative values. Statistical testing (one-way ANOVA, Tukey's HSD) confirmed the absence of significant size effects on the Lyapunov exponents (all $P > 0.1$). This scale independence reveals that mass movements of vastly different magnitudes share similar dynamical attractors despite orders-of-magnitude differences in physical dimensions.

Altogether, this coexistence of divergence (from positive $\lambda_i$) and convergence (from negative $\lambda_i$) within a bounded system (as $\Sigma\lambda_i < 0$) enables local instabilities while maintaining global constraints. These dynamics allow diverse geometric paths to be explored while preserving overall statistical regularity, explaining the consistent statistical properties observed across varied environments.

## 2.3 Cohesion of order and disorder from source to sink

The chaotic dynamics we observed suggest a complex interplay between order and disorder during mass movement. To further explore how order and disorder work together from initiation to deposition, we analyzed changes in entropy from source to sink, which is a measure of disorder that can reveal how predictable or chaotic a geomorphological system becomes over space and time[29,30]. We divided each scar's geometric along-track spatial series (i.e., width, sinuosity, and curvature) partitioned into equal-length parts (i.e., initiation, transport, and deposition zones), then calculated three entropy metrics (i.e., Multiscale Entropy, MSE; Phase Entropy, PhEn; and Kolmogorov-Smirnov entropy, KS) in each zone, to quantify scale-dependent complexity, directional unpredictability, and rates of information loss. Higher entropy values indicate more disordered, unpredictable behavior, while lower values suggest more organized patterns.

Two entropy patterns emerged across small, medium, and large scars (classified by scar area) (Supplementary Tables 4-6; Fig.2f-h; Supplementary Figs.26-30). First, sinuosity and curvature become progressively more disordered downslope. For sinuosity, both MSE and PhEn increased steadily from initiation through transport to deposition. Curvature entropy showed small decrease from initiation to transport, followed by a substantial increase into the deposition zone, particularly for MSE. Second, width entropy follows a distinctive rise-and-fall pattern, typically increased from initiation to peak in the transport zone and then decreased into the deposition zone.

The entropy metrics formed a clear hierarchy: Curvature maintains the highest disorder (entropy values ~1.6 ~ 2.3), where MSE remained nearly constant from initiation (1.639 ± 0.096) through transport (1.625 ± 0.104) to deposition (1.743 ± 0.098), and PhEn similarly showed little change: initiation (2.310 ± 0.206), transport (2.310 ± 0.206), deposition (2.347 ± 0.184). Sinuosity sinuosity shows intermediate disorder that grows with distance (~1.4 ~ 2.1), where MSE rose systematically from initiation (1.355 ± 0.234) through transport (1.411 ± 0.194) to deposition (1.462 ± 0.217), and PhEn followed the same trend: initiation (1.999 ± 0.391), transport (2.088 ± 0.332), deposition (2.143 ± 0.321). Width exhibits the lowest disorder with high variability (~0.6 ~ 1.4), where MSE increased from initiation (0.568 ± 0.553) to transport (0.841 ± 0.489), then decreased during deposition (0.650 ± 0.570), and PhEn showed similar patterns: initiation (1.012 ± 0.877), transport (1.385 ± 0.698), deposition (0.862 ± 0.768).

In sum, curvature and sinuosity become more disordered downslope. Sinuosity entropy increases steadily downslope, reflecting an accumulation of geometric complexity with travel distance[31]. Curvature entropy, however, remains persistently high throughout all stages, indicating near-maximal turning disorder along the path. By contrast, width entropy presents an adaptive process: initially constrained near the source, maximized while negotiating complex terrain, and finally decreasing as energy dissipates during deposition. These patterns reveal how mass movements dynamically balance two fundamental requirements; that is, mass movements must maintain sufficient coherence and momentum for extensive runout yet also possess enough flexibility to navigate uneven terrain and effectively dissipate energy before stopping.

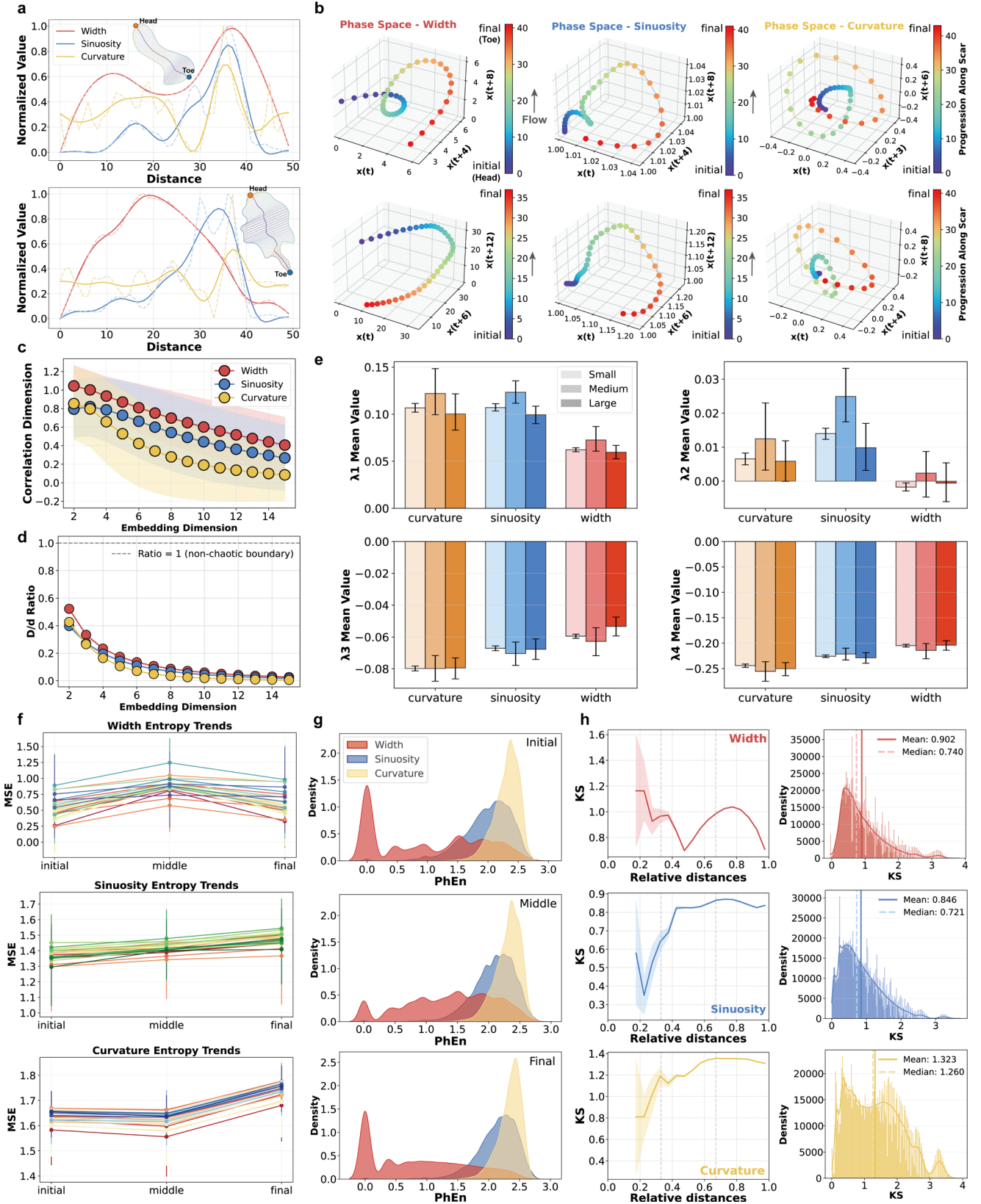

**Fig.2: Deterministic chaos and entropy dynamics in mass movement scar geometry. a,** Representative along-track spatial series of width (red), sinuosity (blue), and curvature (yellow) for two example scars (insets show scars geometry). **b,** Delay-coordinate reconstructions coloured from initiation (blue) to deposition (red) reveal low-dimensional attractors: broad loops for width, intermediate coils for sinuosity and tightly wound spirals for curvature. **c,** Correlation dimension $D$ as a function of embedding dimension $d$. For all signals $D$ saturates well below the identity line (shaded ± s.d.), confirming low-dimensional attractors. **d,** The ratio D/d likewise converges below the non-chaotic boundary (dashed line, $D/d = 1$). **e,** The order of Lyapunov exponents (mean $\pm$ s.d.; $\lambda_1 > \lambda_2 > \lambda_3 > \lambda_4$) for curvature, sinuosity and width, binned by scar-area class (small, medium, large). A positive leading exponent ($\lambda_1$) and a negative sum of exponents across all classes indicate deterministic but dissipative chaos that is independent of scar size. **f,** Multiscale entropy (MSE) averaged over initiation, transport and depositional thirds; thin lines show individual inventories, bold lines the mean $\pm$ s.d. **g,** Phase entropy (PhEn) distributions for all scars, separated by stage (initial, middle, final). These consistently display a clear hierarchy in entropy values: Curvature > Sinuosity > Width, evident at all stages of movement. **h,** Kolmogorov–Smirnov entropy (KS) along the flow path (shaded 95 % confidence) and corresponding histograms: sinuosity and curvature entropy rise downslope, whereas width entropy peaks mid-track then declines.

## 2.4 Long-range persistence demonstrates slow-to-fast hierarchy

Entropy patterns reveal that order and disorder coexist in the spatial variations of geometric signatures along each scar; however, these metrics alone cannot determine whether these trends are purely local or persist downslope. To address this, we quantify long-range correlations by measuring how strongly upstream variations in the three geometric signatures influence subsequent change patterns farther along the movement path.

We first measured autocorrelation functions to confirm how far downstream past states remain influential (Fig.3a,b, Supplementary Fig.31). The decay lags (here first downstream sampling point where the ACF crosses zero) decreased systematically, revealing a slow-to-fast hierarchy: Width maintains influence over the longest distances (mean $\approx$ 11 lags), functions as a slow boundary-setting variable. Sinuosity shows intermediate persistence (mean $\approx$ 7 lags), balancing path continuity with terrain adaptation. Curvature rapidly decorrelates (mean $\approx$ 3 lags), responds rapidly to local perturbations.

To further quantify this persistence hierarchy, we employed four independent methods to estimate the Hurst exponent ($H$) (Fig.3e-i, Supplementary Figs.32-34). (1) Incremental-variance analysis yielded $H = 0.51 \pm 0.16$ for width, indicating persistent behaviour ($H > 0.5$). In contrast, sinuosity ($H = 0.14 \pm 0.27$) and curvature ($H = -0.06 \pm 0.24$), showed anti-persistent characteristics ($H < 0.5$), where trends are more likely to reverse. (2) Periodogram slopes give indistinguishable values for width and sinuosity ($H \approx 1.73 \pm 0.17$ and $1.74 \pm 0.14$; $\beta \approx 2.47 \pm 0.35$ and $2.48 \pm 0.28$) that satisfy the theoretical relation $H = (\beta + 1)/2$, whereas curvature falls in the pink-to-white-noise regime ($H = 0.79 \pm 0.21$; $\beta = 0.59 + 0.41$) (Supplementary Fig.35). (3) Rescaled-range analysis again clusters width (0.819 $\pm$ 0.035) and sinuosity (0.816 $\pm$ 0.054) near $H \approx 0.82$, with curvature lower at 0.693 $\pm$ 0.088. (4) Enhanced detrended-fluctuation analysis (DFA) produces exponents of 1.93 ± 0.21, 1.78 ± 0.15 and 1.12 ± 0.35, respectively (Fig.3b). Curvature's DFA exponent (≈ 1.12) corresponds to a power-spectral slope of approximately 1.0, which is a characteristic of systems near criticality (Supplementary Fig.46-47). Spatial autocorrelation metrics further confirmed these findings: Global Moran's I decreased from width (0.114) to sinuosity (0.056) to curvature (0.051), with corresponding decreases in high-high clustering (11.6%, 2.4%, < 1%), consistent with the expectation that the most rapidly changing parts of a system tend to be localized (Supplementary Figs.38-43). All four lines of evidence converge on a slow-to-fast ordering of width > sinuosity > curvature, consistent with our interpretation of width as a boundary-setting slow variable and curvature as a fast-responding one.

Taken together, width variations exhibit long-range persistence memory, meaning widening or narrowing trends propagate downstream and promote coherent runout. Sinuosity variations, by contrast, show only short-lived persistence. A mass movement path may initially develop multiple bends but eventually straighten out. Curvature variations exhibit strong anti-persistence, with each directional change quickly countered by an opposing turn. This hierarchy reflects mass movements that are dynamically stable enough to reach their final destination while remaining perpetually poised for internal reorganization, consistent with oscillatory correction patterns observed in meandering rivers[32].

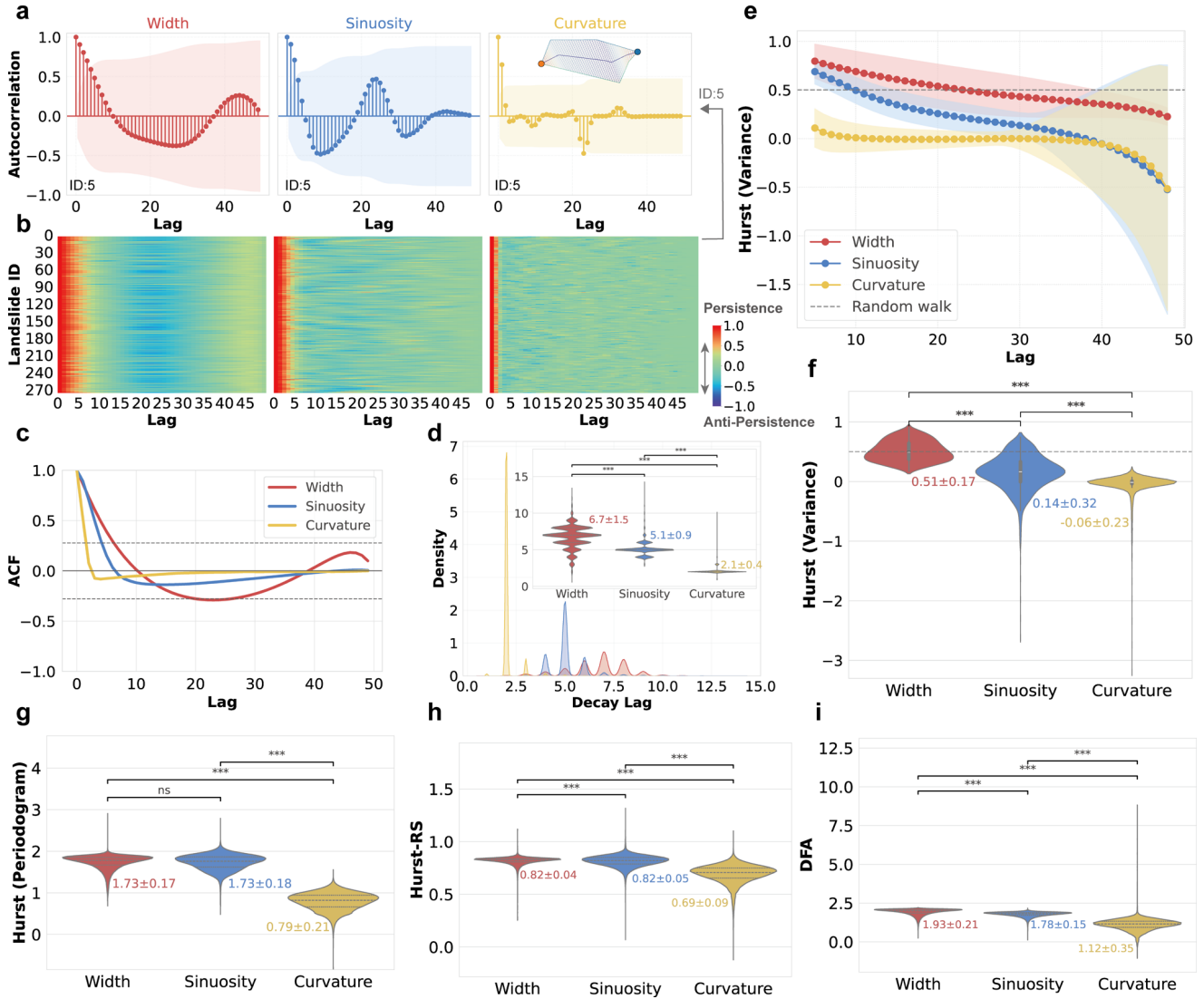

**Fig.3: Spatial memory and persistence hierarchy in mass movement geometric signatures. a,** Example autocorrelation functions (ACFs) for width (red), sinuosity (blue), and curvature (yellow) for an individual mass movement scar (ID:5). Shaded areas represent the 95% confidence interval across all mass movement scars. **b,** Heatmaps of ACFs for width (left), sinuosity (middle), and curvature (right) for mass movement scars from an example inventory (2002CM, n=273 scars shown) (y-axis; individual scar IDs) as a function of lag (x-axis). The colour scale (far right) indicates the autocorrelation coefficient, where red tones (positive values, "Persistence") denote positive correlation and blue tones (negative values, "Anti-Persistence") indicate negative correlation or oscillatory behaviour. **c,** Mean ACFs for width (red), sinuosity (blue), and curvature (yellow) across all mass movement scars, highlighting the differing decay rates. Dashed grey lines indicate zero autocorrelation. **d,** Distributions of decorrelation lags (defined as the first lag where the absolute ACF falls below the two-sided 95 % confidence bound, |ACF| $\ll \pm 1.96/\sqrt{N} \approx 0.25$ for the median path length). Width shows the longest memory (mean ± s.d. = 6.7 ± 1.5), sinuosity intermediate (5.1 ± 0.9) and curvature the shortest (2.1 ± 0.4). **e,** Hurst exponents ($H$) estimated using incremental

variance analysis as a function of lag distance for width (red), sinuosity (blue), and curvature (yellow). Shaded areas are 95% confidence intervals. The dashed grey line at $H = 0.5$ indicates random walk behaviour. **f,** Distributions of Hurst exponents ($H$) derived from incremental variance analysis for width (mean $\pm$ s.d.: 0.51 $\pm$ 0.17), sinuosity (0.14 $\pm$ 0.32), and curvature (-0.06 $\pm$ 0.23). **g,** Distributions of Hurst exponents derived from Periodogram analysis (width: 1.73 $\pm$ 0.17, sinuosity: 1.73 $\pm$ 0.18, curvature: 0.79 $\pm$ 0.21). **h,** Distributions of Hurst exponents derived from Rescaled Range (R/S) analysis (width: 0.82 $\pm$ 0.04, sinuosity: 0.82 $\pm$ 0.05, curvature: 0.69 $\pm$ 0.09). **i,** Distributions of Detrended Fluctuation Analysis (width: 1.93 $\pm$ 0.21, sinuosity: 1.78 $\pm$ 0.15, curvature: 1.12 $\pm$ 0.35). All violin plots show mean $\pm$ s.d. values annotated within the plot area.

## 2.5 Hierarchical organization arising from mutual adjustments

Based on long-range correlations, we found a clear slow-to-fast hierarchy in the three geometric fingerprints of mass movements, confirming that width, sinuosity, and curvature operate at different persistence scales. But can we further quantify this hierarchical organization and capture how these scales interact? We therefore turned to the spectral properties of spatial changes in the geometric signatures, which decompose spatial series into frequency components to reveal scale-dependent hierarchical relationships[33,34,35,36].

Fourier analysis confirmed the hierarchy at the frequency level (Fig.4a,b; Supplementary Fig.44). Specifically, width variations clustered at the longest wavelengths, sinuosity at mid-range, and curvature at the shortest, with dominant periods shortening systematically from width (46.1 $\pm$ 9.6; CV = 0.21) through sinuosity (38.7 $\pm$ 14.2; CV = 0.37) to curvature (18.2 $\pm$ 13.9; CV = 0.76). These differences were statistically significant (Mann-Whitney-Wilcoxon: $p <$ 0.001 for all comparisons). Spectral centroids stepped progressively from low to high ($\approx$ 0.05 → 0.09 → 0.16). Peak counts showed a distinctive pattern: curvature exhibited ~4 significant peaks (mean = 4.41), sinuosity ~1 (mean = 1.15), and width effectively none (mean = 0.37), indicating that curvature has many spatial scales whereas width is dominated by its largest mode. Importantly, only width's spectral properties correlated meaningfully with overall scar geometry, which spectral centroid correlated with both runout distance ($r =$ 0.52) and scar area ($r =$ 0.18), while sinuosity and curvature showed negligible correlations ($r <$ 0.1).

Wavelet analysis showed a similar hierarchy in spectral features (Fig.4c-e, Supplementary Figs. 45-48). Specifically, width energy remained confined to broad scales with minimal variation, sinuosity showed greater spatial variability while favoring larger scales, and curvature exhibited the most heterogeneous signal with episodic high-frequency bursts. Quantitatively, dominant scales decreased systematically from width (43.7 $\pm$ 9.5; CV $\approx$ 0.20) to sinuosity (34.9 $\pm$ 13.8; CV $\approx$ 0.40) to curvature (19.8 $\pm$ 12.5; CV $\approx$ 0.60). Energy ratios confirmed that width concentrated > 80% of its variance in the lowest band, whereas curvature assigned most power to high frequencies (62%). Power-law scaling of wavelet coefficients (width = -0.417, sinuosity = -0.383, curvature = -0.296) demonstrated that fine-scale fluctuations are rapidly damped in width, persist moderately in sinuosity, and remain most pronounced in curvature.

To further quantify such hierarchical organization, we fitted power-spectral densities to a power law (PSD$(f) \propto f^{-\alpha}$), and found a notable pattern in scaling exponents approximating 4-3-2 (Fig.5a-d, Supplementary Figs.49-51). Specifically, width exhibits $a_W = 3.92 \pm 0.33$, resembling fourth-order diffusion processes that suppress fine-scale perturbations while maintaining broad-scale structure. Sinuosity yields $a_S = 2.88 \pm 0.54$, corresponding to third-order dynamics that mediate between momentum conservation and gravitational constraints. Curvature produces $a_C = 1.74 \pm 0.60$, approximating second-order diffusion with rapid responses to local terrain heterogeneities. This exponent ordering ($a_W > a_S > a_C$) persists across different scars size distribution. In width, we observe a moderate decrease in scaling exponent from $3.90 \pm 0.40$ in small scars to $3.01 \pm 0.92$ in large ones ($t = 4.93$, $p < 0.001$, Cohen's $d = 1.26$). Sinuosity and curvature showed similar but weaker trends, with reductions in sinuosity ($2.86 \pm 0.60$ to $2.70 \pm 0.69$; $t = 1.00$, $p = 0.32$, Cohen's $d = 0.25$) and curvature ($1.70 \pm 0.69$ to $1.23 \pm 0.86$; $t = 2.48$, $p = 0.016$, Cohen's $d = 0.61$). The scaling ratio ($a_W : a_S : a_C$) approximates $2.25 : 1.66 : 1$ when normalized, close to the theoretical $2 : 1.5 : 1$ ratio that would be expected for idealized diffusion processes of orders 4, 3, and 2. Furthermore, sensitivity tests that varied detrending (none, linear, quadratic) and resampling (1×, 2×, 4×) consistently reproduced $\alpha$ values of $\approx$ 4, 3 and 2 (Supplementary Figs.55-57).

A critical question emerged: do these hierarchical patterns reflect general interactions across the geometric variables, or merely independent behaviors at different scales? We next employed transfer entropy (TE) to explore this question. After quantifying interactions between width, sinuosity, and curvature, we found asymmetrical leader-follower dynamics in information flow, suggesting directional control cascades that may be related to self-organization processes. From six possible directional couplings, three dominant pathways emerged with significantly higher transfer entropy than their reverse counterparts ($p < 0.001$) (Fig.5e,f; Supplementary Figs.49-51).

First, width-to-curvature coupling exhibited the strongest information transfer (TE = $0.228 \pm 0.044$), exceeding the reverse influence (TE = $0.124 \pm 0.039$). This asymmetry indicates that width predominantly controls subsequent curvature development. Second, width exerts greater influence on sinuosity (TE = $0.176 \pm 0.046$) than sinuosity does on width (TE = $0.143 \pm 0.041$), confirming that lateral expansion constrains potential path sinuosity. Third, sinuosity strongly modulates curvature (TE = $0.239 \pm 0.048$), while curvature exerts a weaker influence on sinuosity (TE = $0.149 \pm 0.043$), suggesting that the extended lateral oscillation of the trajectory (high sinuosity) translates into higher curvature along the path. These directional relationships remained consistent across small, medium, and large mass movements (ANOVA $p < 0.001$).

With these last observations, we are now answer the question of how mass movement self-organize. Specifically, we identified a clear hierarchical organization where width → sinuosity → curvature. This appears to create internal order through a simple mechanism: when a mass movement widens, it gains freedom to develop more complex trajectories, manifesting as increased sinuosity and subsequently higher curvature downslope. We thus conclude that mass movements self-organize into hierarchical dynamical systems similar to other landforms. In these self-organized systems, faster-changing variables become progressively subordinate to slower, larger-scale variables. We observed a similar nested arrangement in mass movements: width represents the slowly evolving variable that establishes boundary conditions for the intermediate variable (sinuosity), which both responds to these constraints and provides context for the fast-responding variable (curvature) that directly adapts to local terrain, forming a universal self-regulating hierarchy.

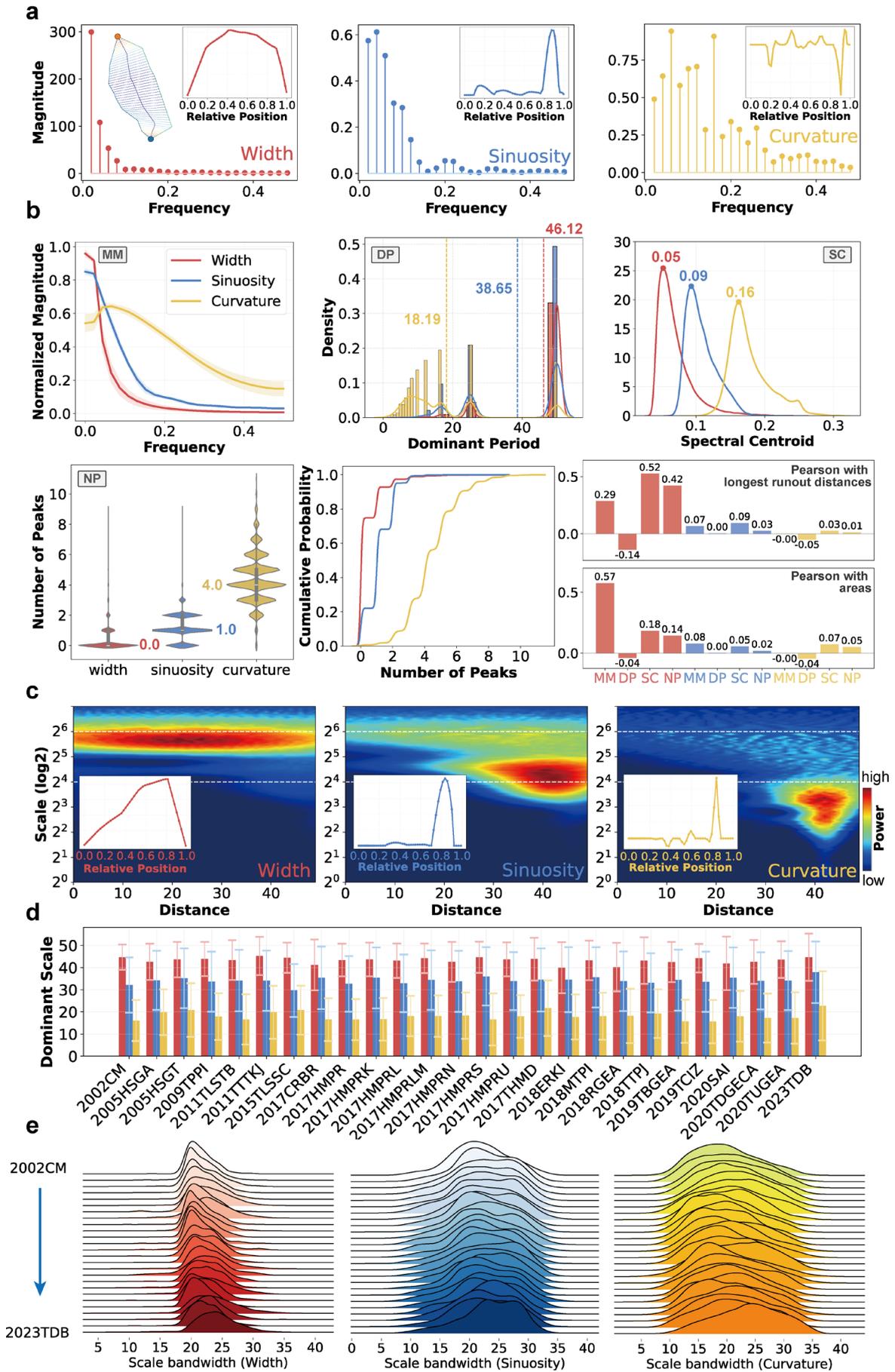

**Fig.4: Fourier and wavelet spectra reveal a three-tier frequency hierarchy in mass movement scar geometry. a,** Fourier-amplitude spectra for a single scar show that width (red) is dominated by low-frequency power, sinuosity (blue) by intermediate bands, and curvature (yellow) by high-frequency energy. Insets give the corresponding spatial profiles. **b,** Statistical analysis of Fourier spectral properties across 65,936 scars. **Mean Normalized Magnitude (MM)** confirms the frequency dominance hierarchy. **Dominant Period (DP)**, distributions of the dominant period shorten markedly from width (mean 46.1) through sinuosity (mean 38.7) to curvature (mean 18.2) (Mann-Whitney-Wilcoxon, $p < 0.001$). **Spectral Centroid (SC)** step from low to high frequency (mean centroids: width $\approx 0.05$, sinuosity $\approx 0.09$, curvature $\approx 0.16$). **Number of Peaks (NP)**, violin and cumulative plots expose a 0-1-4 median peak pattern, indicating rising spectral complexity from width to curvature. Correlation bars (far right) show that width's centroid correlates most strongly with run-out length ($r = 0.52$) and scar area ($r = 0.18$). **c,** Morlet wavelet power for the same scar (log colour scale). Width energy remains at broad scales, sinuosity occupies mid-scales with moderate spatial variability, and curvature exhibits episodic high-frequency bursts. Insets repeat the profiles; dashed white lines mark the dominant scale. **d**, Dominant wavelet scale (mean $\pm$ s.d.) for each inventory; the ordering width > sinuosity > curvature is universal. **e**, Scale bandwidth distributions (ridgeline plots) demonstrate systematic narrowing from slow width fluctuations to rapid curvature adjustments across all inventories. All panel provide provides quantitative evidence for hierarchical frequency structure in mass movements.

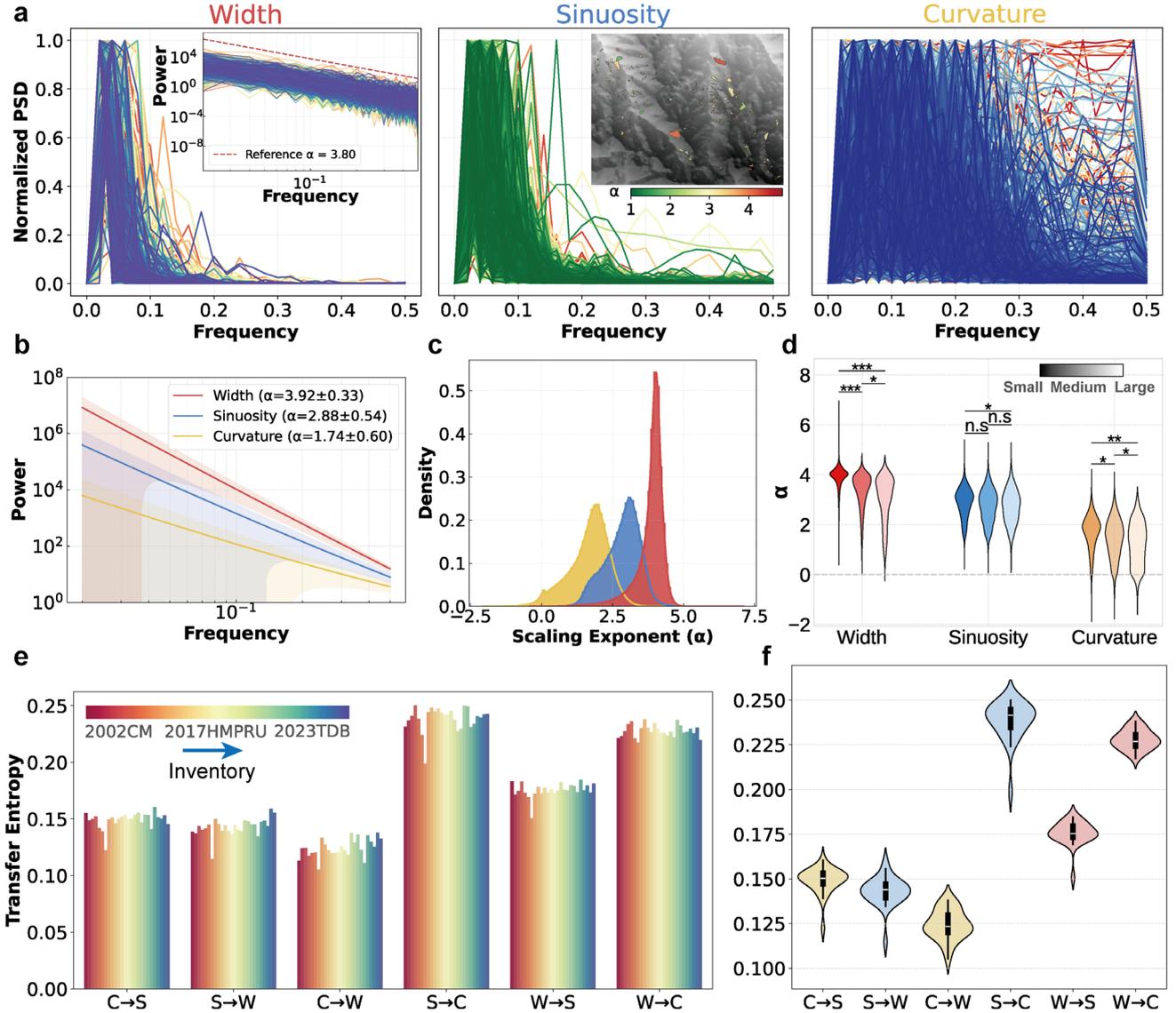

**Fig.5: Power-spectral scaling and information flow reveal hierarchical organization in mass movements. a**, Normalized power-spectral densities (PSDs) for width, sinuosity and curvature of individual scars from inventory 2009TPPI (one line per scar). Insets, PSDs with power-law fits $\text{PSD}(f) \propto f^{-\alpha}$; the width inset shows the inventory-mean slope ($\alpha \approx 3.8$), the sinuosity inset colours scars by $\alpha$ values (hillshade background). **b**, Mean PSDs for 65936 scars (solid lines, shading = $\pm$ 1 s.d. across scars). Mean slopes identify the notable order: width, $\alpha_W = 3.92 \pm 0.33$ ($\approx$ fourth-order); sinuosity, $\alpha_S = 2.88 \pm 0.54$ ($\approx$ third-order); curvature, $\alpha_C = 1.74 \pm 0.60$ ($\approx$ second-order). **c**, Distributions of $\alpha$ show a 4-3-2 order (width $\approx$ 4, sinuosity $\approx$ 3, curvature $\approx$ 2). **d,** Violin plots of $\alpha$ by size class (small, medium, large; grey scale). Width α declines with size, curvature decreases slightly, while sinuosity is size-invariant (n.s.). Significance symbols denote effect size (Cohen's $d$): n.s., $|d| < 0.2$;* $0.2 \leqslant |d| < 0.5$; **, $0.5 \leqslant |d| < 0.8$; ***, $|d| \geqslant 0.8$. **e,** Mean transfer entropy (TE) for the six directional couplings among curvature ($C$), sinuosity ($S$) and width ($W$) across

26 inventories (colour bar). **f,** TE distributions (violins) highlight three dominant pathways: $W \to C$, $W \to S$ and $S \to C$, each significantly stronger than its reverse.

# 3 Discussion

Werner[37] proposed that hierarchical organization governs the complex pattern formation in landforms such as rivers and dunes. In this theory, large-scale forms emerge through local interactions between faster-evolving, smaller-scale components, creating a top-down hierarchy where slow processes set boundary conditions for faster ones[38]. Once higher-level structures emerge, lower-level elements become coordinated by this overarching pattern rather than behaving independently. We found that rainfall-induced mass movements follow similar self-organizing principles.

By treating spatial changes in three geometric signatures (i.e., width, sinuosity, and curvature) as macroscopic fingerprints left by mass movement dynamics, we revealed emergent hierarchical order in mass movements. The spectral analysis identified a distinctive scaling pattern with exponents approximating '4-3-2', leading us to conclude that once a broad flow corridor (width) is established, finer-scale motions (sinuosity and curvature) become orchestrated within it (Fig.6a-c). This emergence of hierarchy, where different geometric components evolve at distinct characteristic scales, mirrors patterns observed in other self-organizing landforms. In meandering rivers, for example, a classic slow-to-fast hierarchy emerges: channel width adjusts slowly to flow conditions and sets constraints, while bending curvature changes rapidly during migration, with wider channels encouraging flow separation and shortcuts that alter the growth of curvature[19,39,40,41]. Our results suggest a similar hierarchical organization operates within rainfall-induced mass movements.

The hierarchical organization opens a question: can the complexity of mass movement self-organization emerge from simple rules? We tested this using a softmax random walk model employing only two physical forces: gravity pulling flows downslope (reflected by topographic slope parameter $\varphi$) and inertia maintaining direction (refected by parameter $\psi$). More details of the softmax random walk model are presented in Supplementary Information Section 5. The results were inspiring: we used multiple parameter configurations and observed emergent paths that consistently matched measured mass movement trajectories, showing that simple rules enable mass movements to navigate complex terrain and find efficient pathways (Fig.6d, Supplementary Figs.58-63). This provides a simple mechanism explaining hierarchical emergence: topographic constraints naturally created 'width-like' boundaries, while the interplay between inertial momentum and gravitational forcing spontaneously organized into 'sinuosity-like' intermediate adjustments and 'curvature-like' local responses.

We suggest that the physical basis for this hierarchical emergence lies in granular physics. Field observations confirm that in disordered systems near yielding thresholds, grain-scale instabilities persistently introduce local unpredictability while bulk constraints maintain global order, with surface roughness serving as a primary control on the transition from bounded to runaway particle motion[42].

This critical balance is maintained through material strength controls that regulate deformation induced by hydraulic pressure gradients, enabling mass movements to preserve large-scale coherence, despite internal instabilities[43,44]. Collective grain interactions continuously generate small-scale randomness and introduce local flexibility, with the resulting high curvature entropy indicating that the system operates near a critical state[45]. At this critical level, transitions between dense, friction-dominated, and dilute, collision-driven regimes create impulsive grain-grain forces and stress fluctuations at the base, causing sharp path deflections that form the curvature and sinuosity peaks we observe. Yet these instabilities are bounded. As we found in our phase-space analysis, low-dimensional attractors impose fundamental constraints on mass movement dynamics, similar to dune morphology, where wind-blown sand self-organizes into limited forms. Similarly, once the large-scale structure (e.g., width) forms, the mass movement follows emergent rules rather than tracking every grain collision, enabling rapid self-adjustment when local perturbations cause directional deviations[46].

All results point to an intriguing organization process, where rainfall-induced mass movements achieve efficient mobility under internal instability through an elegant solution: a nested hierarchy that balances order and chaos. Large-scale structures provide coherence, while constraining disorder, and small-scale adjustments, driven by granular physics, sustain local flexibility, enabling navigation through complex terrain. This balance allows these destructive natural forces to run out over long distances without disintegrating. The recognition that such complex, destructive phenomena follow predictable organizational rules opens exciting possibilities for improving hazard assessment and risk reduction.

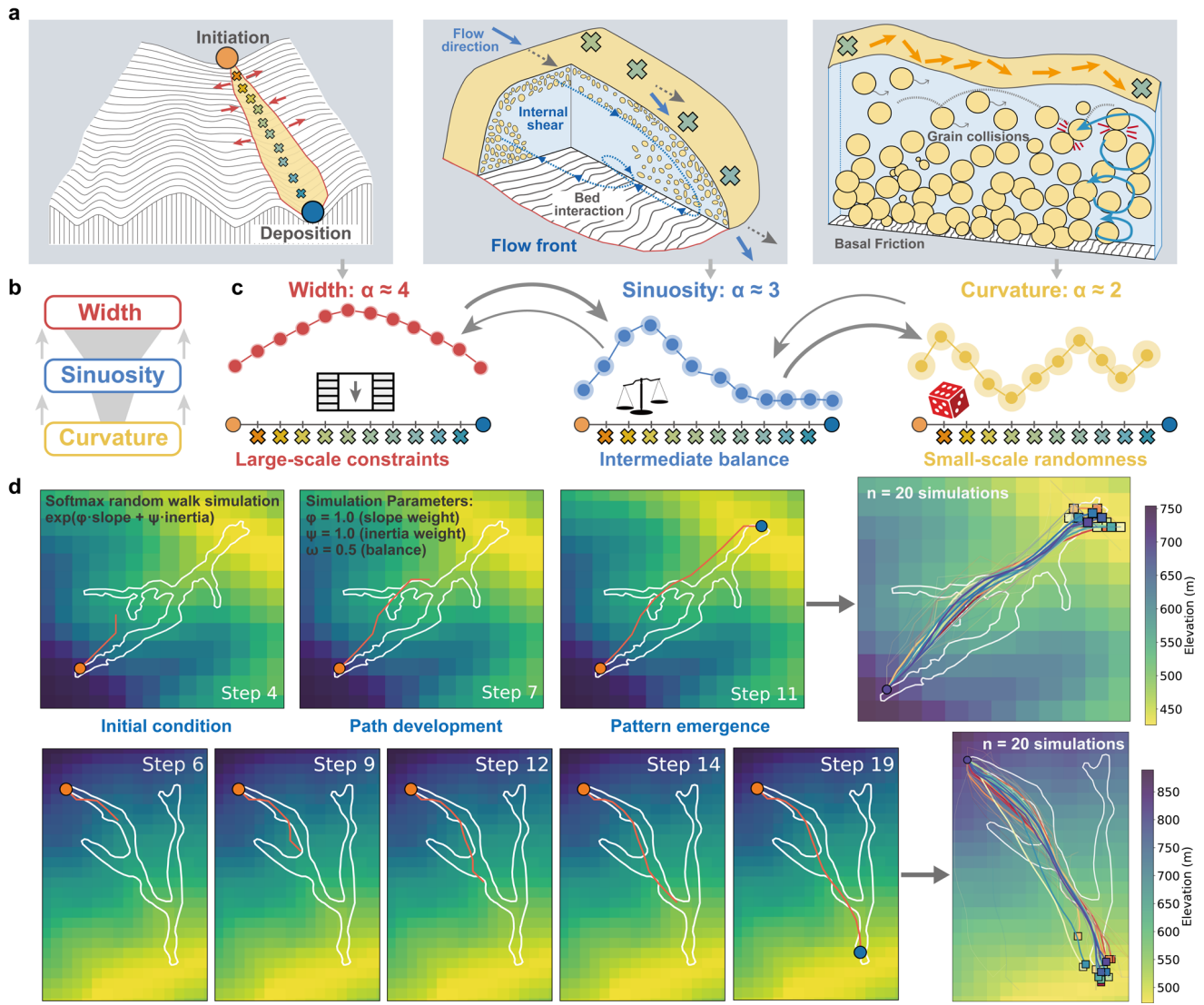

**Fig.6: Hierarchical self-organization in rainfall-induced mass movements. a**, Conceptual sketch of the physical processes that shape the three principal geometric signatures of a scar. Width (left) is set by broad-scale mobilisation and lateral spreading from initiation (orange circle) to deposition (blue circle). Sinuosity (centre) arises from internal shear, bed interaction and flow-front dynamics that deflect the path from a straight line. Curvature (right) reflects grain-scale collisions, turbulent eddies and reactions to basal friction and micro-topography, generating rapid local directional changes. **b**, Empirically inferred control hierarchy: path width acts as the primary (slow) variable, conditioning sinuosity, which in turn modulates curvature. **c**, Spectral-scaling hierarchy derived from 65936 scars. Width evolves with a steep PSD exponent $\alpha_W \approx 4$ (≈fourth-order diffusion), sinuosity with $\alpha_S \approx 3$ (≈third-order), and curvature with $\alpha_C \approx 2$ (≈second-order), indicating progressively faster-and more locally responsive-dynamics from W to S to C. **d**, Conceptual terrain–inertia model implemented as a softmax random walk (see Supplementary Information Section 5). A moving mass steps across the DEM with transition probability $P((x,y) \to (i,j)) \propto \exp[(1-\omega)\psi I_{ij} + \omega\phi S_{ij}]$, where $S_{ij}$ is downslope bias, $I_{ij}$ an inertia term, and $\phi$, $\psi$, $\omega$ are weighting parameters. Left, sequential positions for one

realisation (orange path; head, orange circle; toe, blue circle) illustrate how a coherent trajectory emerges from purely local rules. Right, ensembles of 20 runs with varied ($\phi$, $\psi$, $\omega$) show a consistent run-out envelope and characteristic morphology, demonstrating that large-scale path geometry is robust to stochastic variability yet sensitive to the relative weighting of slope and inertia.

# 4 Methods

## 4.1 Mass movement scar dataset

We compiled a global dataset of 26 rainfall-induced mass movement inventories (2002 ~ 2023), documenting events triggered by extreme rainfall across diverse geographic and geological settings. Events included tropical cyclones (Micronesia, 2002; Philippines, 2009), hurricanes (Puerto Rico and Dominica, 2017), typhoons (Japan, 2011; Philippines, 2018; Beijing, 2023), convective storms (Brazil, 2011; Colombia, 2015), and intense rainfall (Brazil, 2017). Most inventories were validated using field investigations and remote sensing analyses, ensuring the inclusion of only rainfall-triggered mass movements[47,48,49,50,51,52,53,54]. Terrain morphology was derived from a multi-temporal Digital Elevation Model (DEM) database comprising SRTM (30 m resolution, 2000), ALOS PALSAR DEM (12.5 m/30 m resolution, 2006 ~ 2011), and Copernicus DEM (30 m resolution, 2021). We prioritized pre-event DEMs to establish a consistent topographic baseline across the 2002 ~ 2023 dataset, thereby minimizing the effects of post-event modification.

We then preprocessed polygon geometries by converting multi-polygon features to single components, retaining the largest area polygon from each feature. We applied adaptive smoothing based on geometric complexity, quantified as the ratio of the polygon perimeter to the circumference of an equivalent-area circle. Polygons exceeding a complexity threshold of 1.5 underwent iterative midpoint smoothing (maximum 15 iterations) while preserving vertices with angles < 20° to retain critical geometric features.

## 4.2 Geometric signatures calculation

Inspiring from the approaches for analyzing river geometries, we employed three signatures to characterize mass movement dynamics: (1) width ($W$) quantifying lateral variation, (2) sinuosity ($S$) measuring path deviation, and (3) curvature ($C$) capturing directional changes along transport paths [55,56,57,58].

Our automated extraction approach combined computational geometry with elevation analysis. For each scar polygon, we identified head and toe points by finding the maximum boundary separation distance, then used DEM-derived elevations to classify them (higher elevation = head, lower = toe), establishing consistent source-to-sink orientation. Lateral boundaries were delineated relative to the downslope direction, and each trajectory was resampled to 100 equidistant points using linear interpolation.

Width measurements used geodesic calculations between corresponding boundary point pairs at equivalent positions along each path. We connected points at the same relative distance from the head along each boundary, creating cross-sectional transects that captured width variations along the scar. Each measurement received a normalized position value (0 at head, 1 at toe) based on geodesic distance from the head point, ensuring comparable width profiles across different scar sizes.

Centerlines were generated by connecting width measurement midpoints, creating flow paths that captured mass movement trajectory variations. This paired-point approach showing consistency with long main path distances derived from Voronoi-based methods (see Supplementary Information Section 1). Based on these centerlines, we then calculated sinuosity ($S$) using a moving window approach. For each point $i$ along the centerline, we extracted a local segment of up to ten neighboring points-five on either side of i when available (window size $n = 10$), truncating at path boundaries if fewer than five points existed upstream or downstream. Let the UTM projected coordinates of the window be $\{(x_1, y_1), ..., (x_m, y_m)\}$, where $m \leq 10 + 1$, we defined the incremental sinuosity $S_i$ following river geometry analysis methods as:

$$S_i = \frac{L_i}{D_i} \quad (1)$$

where $L_i$ is the total polyline length in the window:

$$L_i = \sum_{k=1}^{m-1} \sqrt{(x_{k+1} - x_k)^2 + (y_{k+1} - y_k)^2} \quad (2)$$

$D_i$ is the straight-line (Euclidean) distance between the first and last points of the window,

$$D_i = \sqrt{(x_m - x_1)^2 + (y_m - y_1)^2} \quad (3)$$

A spatial series of $S_i$ values was computed at each point along the path using 50 sample point, as indicator for capturing local path tortuosity: values near 1.0 indicate straight segments, while higher values reflect increasing path deviation from direct downslope flow.

We then calculated curvature ($C$) along extracted centerlines at each point using differential geometry, to observe how mass movements adjust their paths in response to terrain constraints, capturing patterns of local directional changes from initiation to deposition:

$$C = \frac{x'y'' - y'x''}{(x'^2 + y'^2)^{3/2}} \quad (4)$$

where $x'$ and $y'$ denote first derivatives, and $x''$ and $y''$ represent second derivatives of the $x$ and $y$ coordinates with respect to distance along the path. The established head-to-toe directionality of mass movements enabled us to interpret curvature signs consistently. Positive curvature values indicate rightward turns while negative values represent leftward turns relative to downslope progression. We verified movement directionality by comparing distances from each centerline endpoint to the identified head and toe points, reordering coordinates when necessary to ensure consistent orientation. Since curvature calculations involve second-order spatial derivatives that amplify small fluctuations, we applied Gaussian smoothing and robust outlier detection to reduce noise before feature extraction.

To decode the organizational processes governing mass movements, we transformed the three extracted geometric signatures into dynamic fingerprints[59]. For each scar, after extracting width, sinuosity, and curvature along its centerline, we sampled these measurements at 50 equidistant points. Normalizing the position coordinates from 0 (head) to 1 (toe) created a standardized spatial series. These standardized profiles serve as dynamic fingerprints, systematically capturing how each signature evolves from source to sink and providing the basis for analyzing the organizational rules governing mass movement.

## 4.3 Self-organization analysis

To investigate organizational rules in mass movement organization, we treated along-track variations in geometric signatures as spatial signals and analyzed five characteristics of self-organizing systems: (i) deterministic chaos, (ii) entropy change, (iii) long-range correlations, (iv) spectral properties, and (v) information flow [60,61,62].

**Deterministic chaos analysis.** We reconstructed phase space dynamics using Takens' embedding theorem with embedding dimension $d = 10$. We first applied low-pass filtering to reduce noise, then optimized delay parameters by computing mutual information functions and selecting delays at the first minimum. We calculated correlation dimensions ($D$) using the Grassberger-Procaccia algorithm by computing correlation integrals $C(\varepsilon)$ at increasing distance scales ε and fitting power-law relationships $C(\varepsilon) \sim \varepsilon^D$. Lyapunov exponent spectra were calculated using Eckmann's algorithm, by tracking the evolution of initially orthogonal vectors in phase space and quantifying the exponential rates at which nearby trajectories diverge.

**Entropy change analysis.** We characterized entropy changes from source to sink by dividing each path into initiation, transport, and deposition thirds, then computing three entropy measures for each segment. For Phase Entropy (PhEn), we calculated phase angles between successive path segments, constructed second-order difference plots, and computed probability distributions of directional changes. For Multiscale Entropy (MSE), we first calculated Incremental Entropy at the finest scale, then constructed coarse-grained series by averaging non-overlapping windows and recalculated entropy at each scale. For Kolmogorov-Smirnov entropy, we embedded geometric series into high-dimensional vectors and applied rolling windows to track information emergence rates along flow paths.

**Long-range correlation analysis.** We investigated the long-range correlation using autocorrelation functions and Hurst exponents. Autocorrelation functions quantify how strongly morphological features at one point correlate with features at progressively longer distances downstream. We first computed autocorrelation coefficients at increasing lag distances for each geometric signature, then identified decorrelation lengths where correlations drop below significance thresholds. To distinguish persistent from anti-persistent patterns, we estimated Hurst exponents through four complementary approaches. For aggregated variance analysis, we calculated variance of incremental $(X_{i+\Delta x} - X_i)$ differences at increasing lag intervals $\Delta x$ and fitted power-law relationships to determine scaling exponents; for periodogram analysis, we computed power spectral densities and derived Hurst exponents $\beta$ from spectral slope relationships $H = (1 - \beta)/2$; for rescaled range analysis, we divided each series into segments of varying length, computed rescaled range statistics R/S, and fitted scaling relationships; for enhanced detrended fluctuation analysis, we removed polynomial trends from integrated series and analyzed fluctuation scaling at multiple scales.

**Spectral analysis.** We applied Fourier Transforms by computing amplitude spectra to transform each spatial series into the frequency domain and then identify characteristic wavelengths within each dynamic fingerprint[63,64]. For continuous wavelet analysis, we convolved each series with Morlet wavelets (cmor 1.5 ~ 1.0) at 128 logarithmically spaced scales ($2^0$ ~ $2^7$), computing wavelet coefficients that preserve both frequency and spatial information. We extracted dominant scales from wavelet power spectra and calculated energy ratios across frequency bands to quantify multi-scale organization.

**Information flow analysis.** We employed transfer entropy by first normalizing all spatial series of geometric signatures, then discretizing values using adaptive quantile binning (5 bins). We computed conditional probabilities for all possible state transitions and calculated transfer entropy (TE) for each directional pair, and applied light Gaussian smoothing to reduce discretization artifacts[65].

## 4.4 Terrain–inertia trade-offs model

To reproduce organizational rules in mass movement systems, we adapted a terrain-driven stochastic model based on methods used for simulating river avulsion pathways[66]. To model simple rules, our approach considers only topographic gradients and movement inertia within a probabilistic framework using a physics-informed softmax random walk algorithm[67].

The model calculates the probability of moving from current grid cell $(x, y)$ to neighboring cell $(i, j)$ by blending topographic slope $(S_{i,j})$ and inertia $(I_{i,j})$. Topographic slope represents elevation differences: $S_{i,j} = (e_c - e_n)/d_{i,j}$, where $e_c$ and $e_n$ is elevation and $d_{i,j}$ is Euclidean distance. Movement inertia quantifies alignment between previous movement direction and potential next moves using cosine similarity, scaled to [0, 1] to ensure higher values indicate greater directional alignment.

These factors are combined using weighting parameters $\varphi$ (slope influence), $\psi$ (inertia influence), and blending factor $\omega$. Combined weights are transformed into movement probabilities using the softmax function, ensuring probabilities sum to 1 while excluding moves returning to the immediate last position to prevent oscillations.

At each iteration, the algorithm calculates slope and inertia values for valid neighbors, computes combined weights, applies softmax transformation, and randomly selects the next position based on computed probabilities. Simulations continue until reaching predefined stopping criteria (minimum slope, maximum path length, or grid boundary exit).

This approach enables the investigation of how terrain gradients and inertial effects influence trajectory development from simple, local rules, without requiring detailed geotechnical parameters. For more details, please refer to the Supplementary Information Section 5.

# Method references

# Supplementary information.

Additional details on several aspects of this work, including supplementary figures, the proposed data-driven framework for capturing organizational rules from mass movement scar fingerprints, and the terrain-inertia model for validating mass movement organizational rules, are available in the supplementary information.

# Extended Data Figures

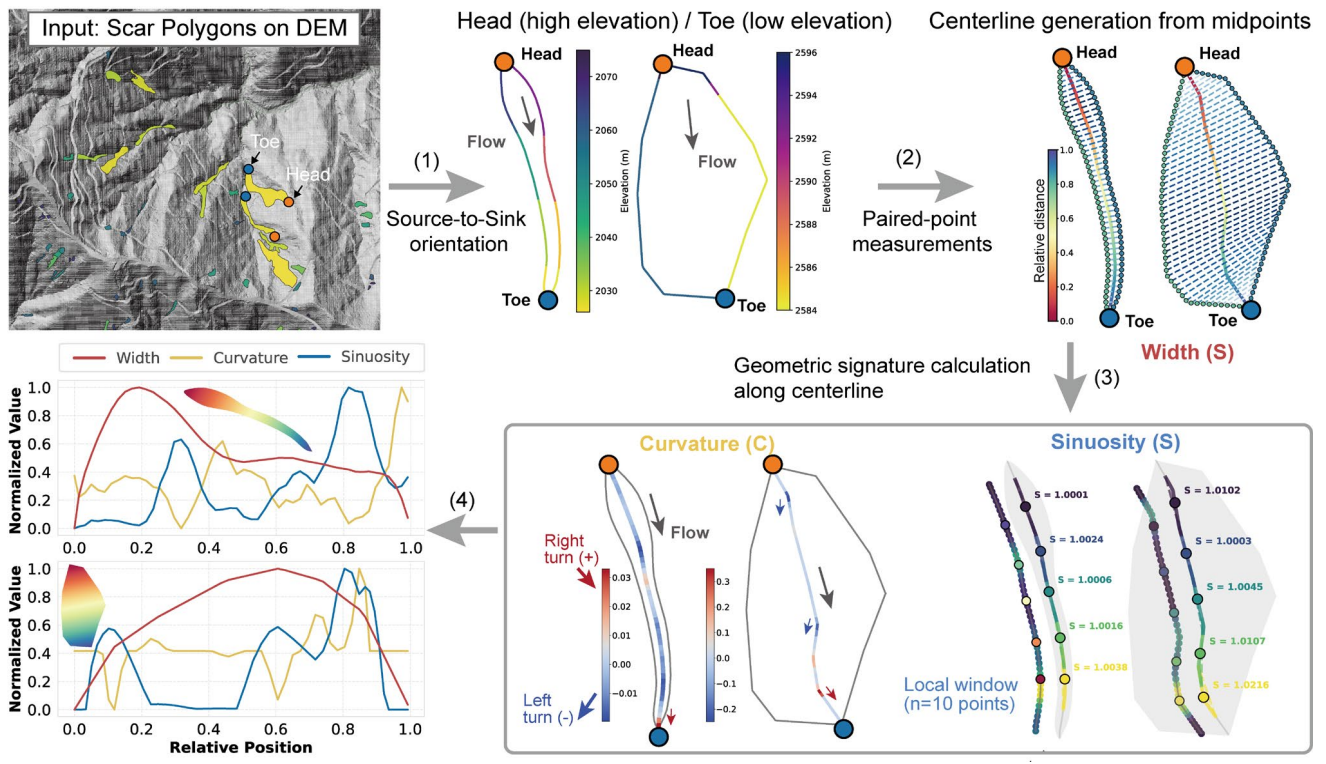

**Extended Data Fig.1 Extraction and calculation of mass movement scars geometric signatures.** The workflow illustrates the following steps: (1) Identification of head and toe points from the mass movement scar polygons on a DEM. (2) Generation of scar centerlines from the midpoints of paired-point width (W) measurements taken at 100 equidistant points along the scar boundaries. (3) Calculation of local curvature (C), representing the rate of directional change (positive for right turns, negative for left turns relative to flow), and incremental sinuosity (S), representing path deviation within a 10-point moving window along the centerline. (4) Generation of normalized along-track profiles for W, C, and S against relative position (0=Head, 1=Toe).

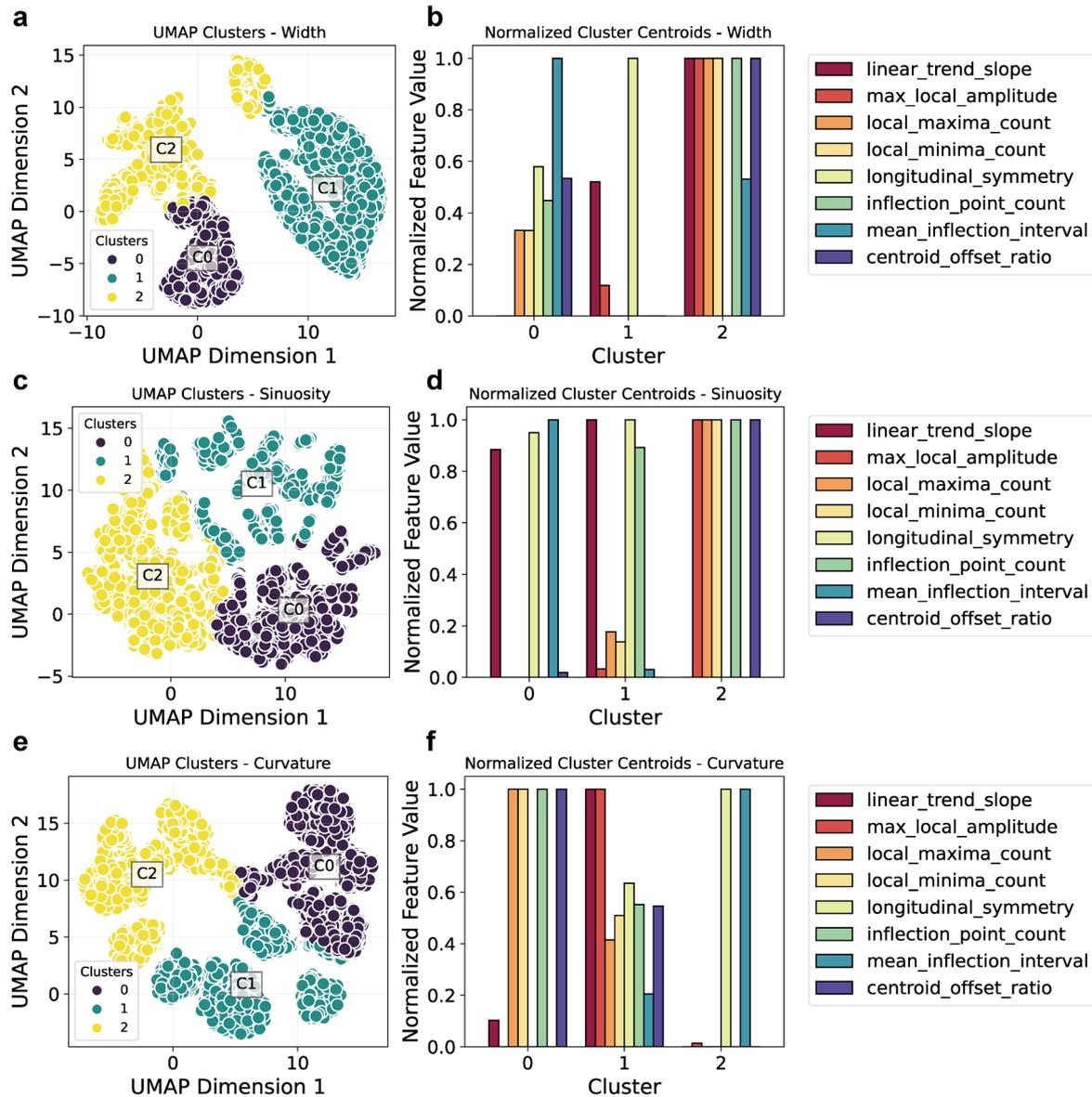

**Extended Data Fig.2 Universal geometric patterns of mass movement scars identified by manifold learning.** **a**, **c**, **e**, UMAP embeddings of all scars based on derived geometric attributes for width (a), sinuosity (c), and curvature (e). Colours denote distinct geometric archetypes (C0, C1, C2) identified through clustering. **b**, **d**, **f**, Normalized feature values for the centroids of each archetype identified in **a**, **c**, **e** respectively. For width (archetypes from **a**, centroids in **b**), archetypes represent (C0) constricted / tapering, (C1) highly symmetric / uniform, and (C2) pronouncedly oscillatory profiles. For sinuosity (archetypes from **c**, centroids in **d**), archetypes range from (C0) low / stable, to (C1) trend-dominated / asymmetric, and (C2) highly oscillatory paths. For curvature (archetypes from **e**, centroids in **f**), archetypes are (C0) frequent / sharp turns, (C1) trend-dominated/asymmetric sequences, and (C2) symmetric profiles with fewer / broader turns. The consistent emergence of these archetypes (mean silhouette scores: width = 0.598, sinuosity = 0.421, curvature = 0.453) and high centroid similarity across inventories (mean similarity: Width = 0.955, Sinuosity = 0.993, Curvature = 0.899) indicate fundamental underlying principles in mass movement planform development.

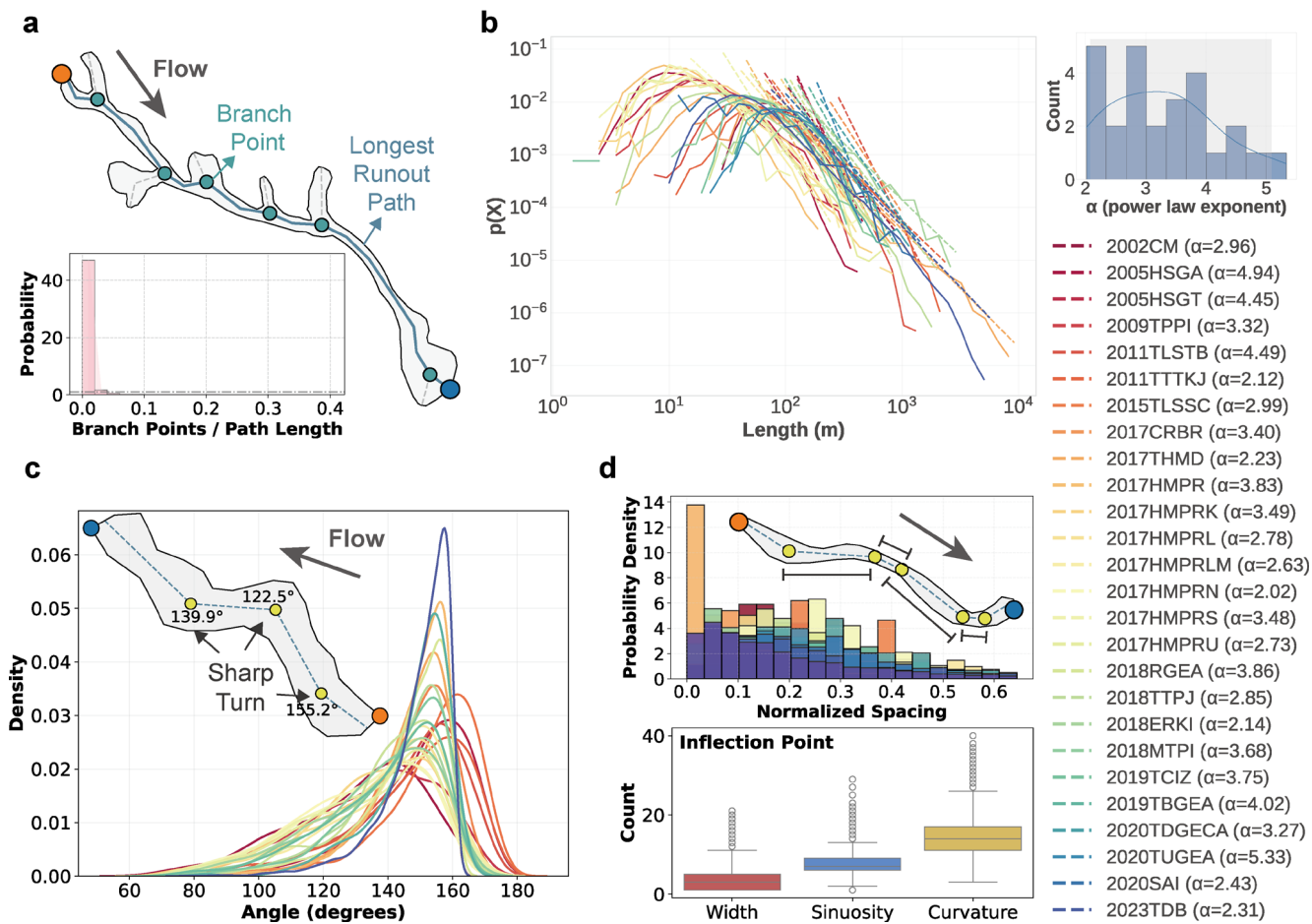

**Extended Data Fig.3 Universal geometric patterns from Voronoi-based scar skeleton analysis.** Voronoi-derived skeletons from 65,936 mass movement scars reveal universal topological features, corroborating planform regularities observed from full scar geometries. **a,** Conceptual skeleton illustrating the longest runout path and branch points. Inset: Normalized branch point locations along the longest path exhibit a consistent upper bound near 0.4. **b,** Longest runout path lengths for 26 inventories follow heavy-tailed, power-law distributions, with fitted exponents α consistently clustering around a mean of 3.3. **c** Sharp turns (angular deviation > 75°) on skeletons (conceptual example shown) predominantly exhibit deviation angles between 140° and 160° (mean 144.6° ± 0.07° s.d.). **d,** Top: Consecutive sharp turns are closely spaced (mean normalized spacing 0.2632 ± 0.1667 s.d.; conceptual illustration and distribution shown). Bottom: Box plots show systematically increasing inflection point counts from full planform geometry analyses of width (mean 3.27 ± 2.60), sinuosity (mean 7.34 ± 1.96), to curvature (mean 14 ± 6), indicating geometric adjustments intensify at smaller scales.

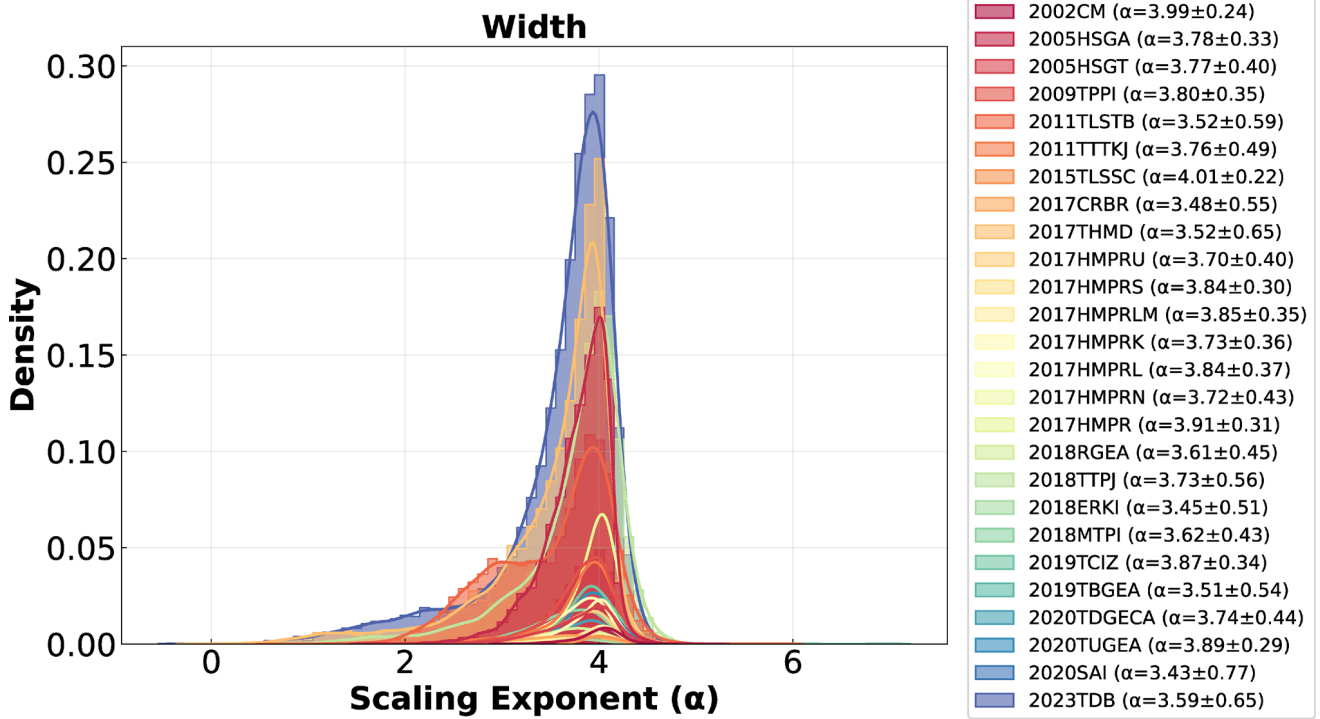

**Extended Data Fig.4 Width spectra converge on a universal fourth-order scaling law.** Kernel density plots of scaling exponents ($a$) from power-law fits to width power spectral densities ($PSD(f) \propto f^{-\alpha}$) across 26 global rainfall-triggered inventories (one curve per inventory; legend gives mean ± s.d.). Width consistently exhibits scaling exponents clustering around $a \approx 4$ (global mean: 3.92 ± 0.33), characteristic of fourth-order diffusion processes that suppress fine-scale perturbations while maintaining broad-scale structure. The remarkable convergence across diverse geographic, climatic, and geological settings demonstrates universal scaling behavior independent of local environmental conditions. Individual inventory means range from 3.45 to 4.01, with most distributions tightly clustered around the theoretical fourth-order value. This scaling signature reflects width's role as the primary boundary-setting variable that establishes large-scale flow corridors through slow-evolving, persistent geometric adjustments. The consistency of fourth-order scaling across all inventories supports the interpretation that width functions as the dominant organizing variable in the self-organized hierarchical structure of mass movements, representing the slow-evolving component that constrains faster geometric variables in the hierarchical organization.

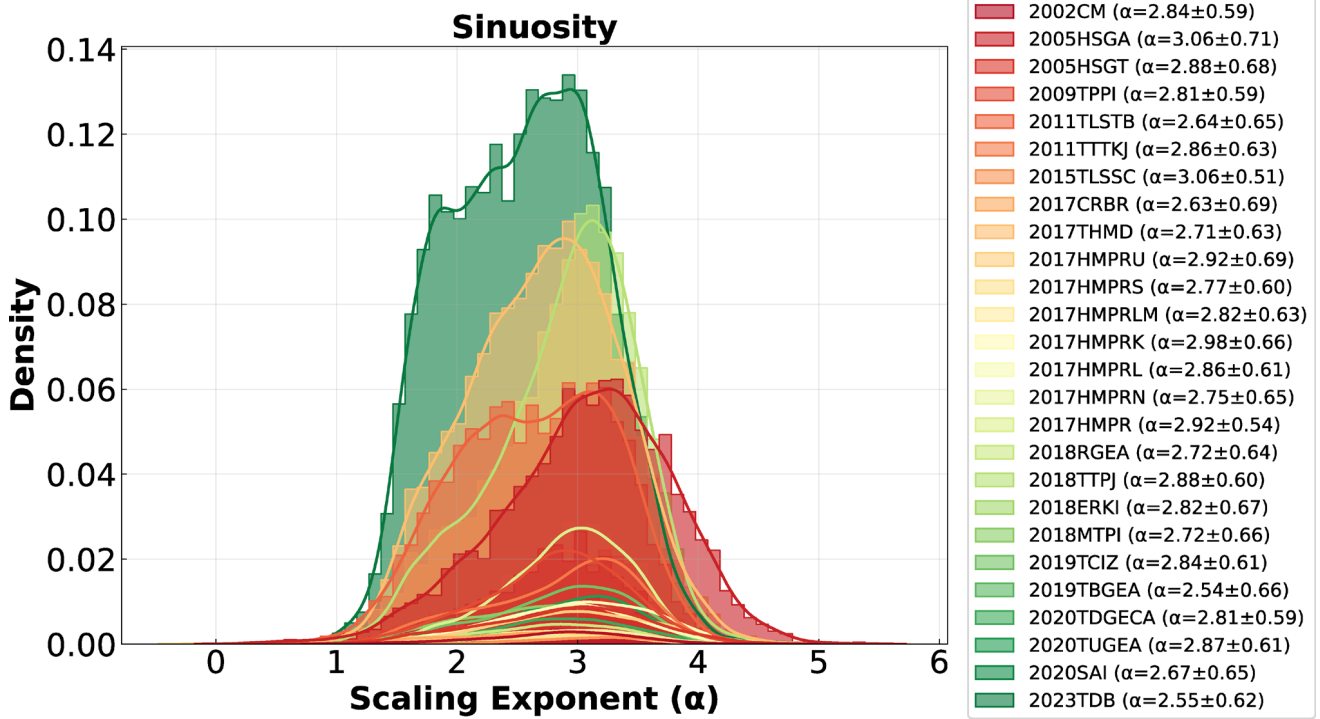

**Extended Data Fig.5 Sinuosity spectra converge on a universal three-order scaling law.** Kernel density plots of scaling exponents ($a$) from power-law fits to sinuosity power spectral densities ($PSD(f) \propto f^{-\alpha}$) across 26 global rainfall-triggered inventories (one curve per inventory; legend gives mean ± s.d.). Sinuosity consistently exhibits scaling exponents clustering around $a \approx 3$ (global mean: 2.88 ± 0.54), characteristic of third-order dynamics that mediate between momentum conservation and gravitational constraints. The convergence across diverse geographic, climatic, and geological settings demonstrates universal scaling behavior independent of local environmental conditions. Individual inventory means range from 2.54 to 3.06, with distributions centered around the theoretical third-order value. This scaling signature reflects sinuosity's role as an intermediate-scale variable that balances persistence with terrain adaptation, operating subordinate to width's boundary-setting constraints while modulating curvature responses. The consistency of third-order scaling across all inventories supports the interpretation that sinuosity functions as the intermediate organizing variable in the self-organized hierarchical structure of mass movements, representing the component that mediates between slow width evolution and rapid curvature adjustments in the nested temporal organization.

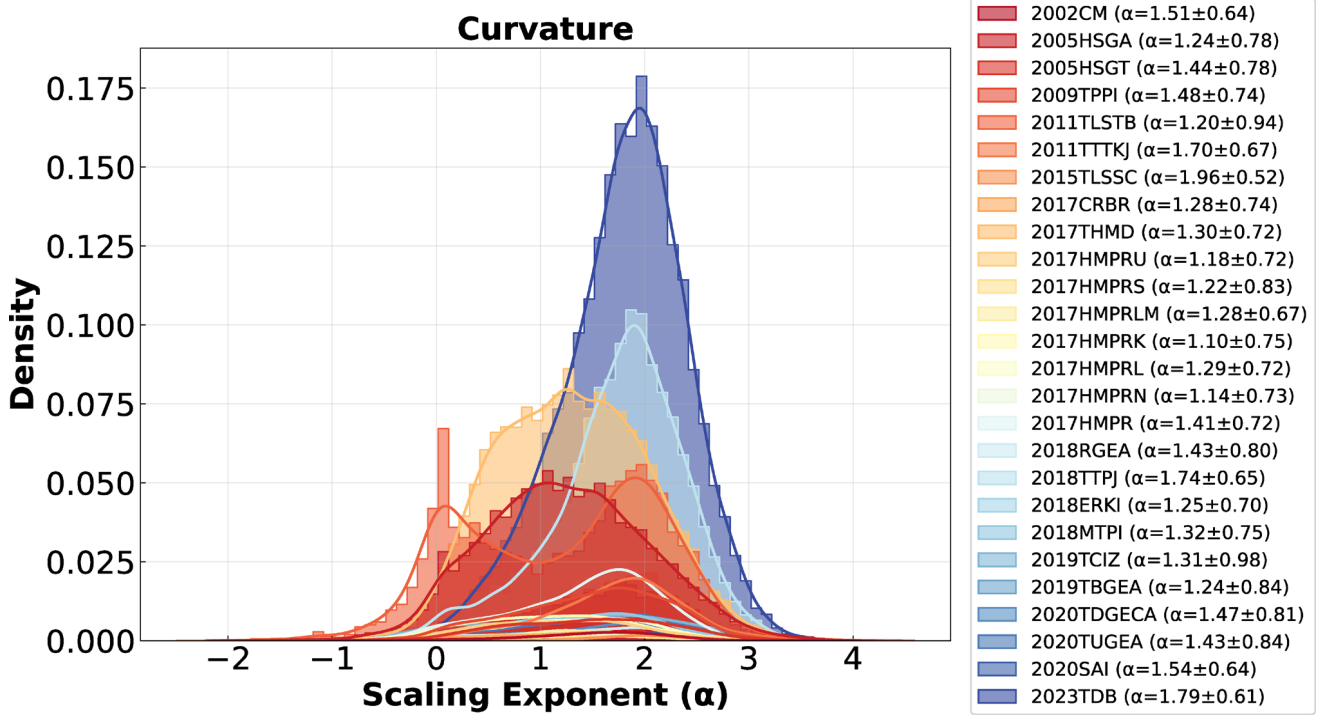

**Extended Data Fig.6 Curvature spectra converge on a universal two-order scaling law.** Kernel density plots of scaling exponents ($a$) from power-law fits to curvature power spectral densities ($PSD(f) \propto f^{-\alpha}$) across 26 global rainfall-triggered inventories (one curve per inventory; legend gives mean ± s.d.). Curvature consistently exhibits scaling exponents clustering around $a \approx 2$ (global mean: $1.74 \pm 0.60$), characteristic of second-order diffusion with rapid responses to local terrain heterogeneities. The convergence across diverse geographic, climatic, and geological settings demonstrates universal scaling behavior independent of local environmental conditions. Individual inventory means range from 1.10 to 1.96, with distributions centered around the theoretical second-order value. This scaling signature reflects curvature's role as the fastest-responding variable that provides immediate adaptation to micro-topographic variations while operating within constraints established by width and sinuosity. The consistency of second-order scaling across all inventories supports the interpretation that curvature functions as the most localized component in the self-organized hierarchical structure of mass movements, representing the rapidly-adjusting variable that enables instantaneous terrain negotiation within the nested temporal organization established by slower geometric variables.